\documentclass[sigconf]{acmart}

\AtBeginDocument{%
  }

\copyrightyear{2026}
\acmYear{2026}
\setcopyright{rightsretained}
\acmConference[ICSE '26]{2026 IEEE/ACM 48th International Conference on Software Engineering}{April 12--18, 2026}{Rio de Janeiro, Brazil}
\acmBooktitle{2026 IEEE/ACM 48th International Conference on Software Engineering (ICSE '26), April 12--18, 2026, Rio de Janeiro, Brazil}
\acmPrice{}
\acmDOI{10.1145/3744916.3764557}
\acmISBN{979-8-4007-2025-3/26/04}
\usepackage{xcolor}
\usepackage{framed}
\usepackage{amsmath}
\usepackage{graphicx}
\usepackage{multirow}

\begin{document}
\renewenvironment{leftbar}{%
  \def\FrameCommand{%
    {\color{black!75}\vrule width 1pt} 
    \hspace{0pt} 
  }%
  \MakeFramed{\advance\hsize-\width \FrameRestore}%
}%
{\endMakeFramed}

\newenvironment{findingbox}[1]{%
    \begin{leftbar}%
    \noindent\colorbox{black!75}{\makebox[\dimexpr\linewidth-8pt\relax][l]{\textcolor{white}{\bfseries #1}}}%
    \vspace{4pt}\newline\noindent\ignorespaces%
    \color{black}%
}{%
    \end{leftbar}%
}
\title{SeRe: A Security-Related Code Review Dataset Aligned with Real-World Review Activities}

\author{Zixiao Zhao}
\affiliation{%
  \institution{Key Lab of HCST (PKU), MOE; SCS}
  \institution{Peking University}
  \city{Beijing}
  \country{China}
}
\email{2301111987@stu.pku.edu.cn}

\author{Yanjie Jiang}
\authornote{Corresponding author.}
\affiliation{%
  \institution{Peking University}
  \institution{College of Intelligence and Computing, Tianjin University}
  \country{}
}
\email{yanjiejiang@tju.edu.cn}

\author{Hui Liu}
\affiliation{%
  \institution{School of Computer Science \& Technology}
  \institution{Beijing Institute of
 Technology}
 \country{}
}
\email{liuhui08@bit.edu.cn}

\author{Kui Liu}
\affiliation{%
  \institution{Huawei Software Engineering Application Technology Lab}
  \country{China}
}
\email{brucekuiliu@gmail.com}

\author{Lu Zhang}
\affiliation{%
  \institution{Key Lab of HCST (PKU), MOE; SCS}
  \institution{Peking University}
  \city{Beijing}
  \country{China}
}
\email{zhanglucs@pku.edu.cn}

\renewcommand{\shortauthors}{Trovato et al.}

\begin{abstract}

Software security vulnerabilities can lead to severe consequences, making early detection essential. Although code review serves as a critical defense mechanism against security flaws, relevant feedback remains scarce due to limited attention to security issues or a lack of expertise among reviewers. Existing datasets and studies primarily focus on general-purpose code review comments, either lacking security-specific annotations or being too limited in scale to support large-scale research. To bridge this gap, we introduce \textbf{SeRe}, a \textbf{security-related code review dataset}, constructed using an active learning-based ensemble classification approach. The proposed approach iteratively refines model predictions through human annotations, achieving high precision while maintaining reasonable recall. Using the fine-tuned ensemble classifier, we extracted 6,732 security-related reviews from 373,824 raw review instances, ensuring representativeness across multiple programming languages. Statistical analysis indicates that SeRe generally \textbf{aligns with real-world security-related review distribution}. To assess both the utility of SeRe and the effectiveness of existing code review comment generation approaches, we benchmark state-of-the-art approaches on security-related feedback generation. By releasing SeRe along with our benchmark results, we aim to advance research in automated security-focused code review and contribute to the development of more effective secure software engineering practices.

\end{abstract}



\maketitle

\section{Introduction}
Security is a fundamental concern in modern software development, as security vulnerabilities can lead to severe financial and operational consequences \cite{10.5555/1349703}. As late-stage fixes are often costly and complex, it is crucial to address security issues early in the development process, for example, at code review time. Code review, a widely adopted practice in software development, plays a critical role in improving software quality by identifying defects, enforcing coding standards, and facilitating knowledge sharing \cite{yang2024survey}. During this process, reviewers raise issues in the code and engage in discussions with developers, who then refine their code changes based on the feedback.  As an incremental and continuous process, code review allows for the early detection of potential security flaws at a relatively low cost, making it an effective approach to mitigating security risks. Research by Thompson et al. \cite{thompson2017large} demonstrates a negative correlation between code review coverage and the prevalence of security vulnerabilities in software projects, highlighting substantial impact of reviews in enhancing code security.

In practice, however, security-related comments account for only a small fraction of code review feedback. For example, Di Biase et al.~\cite{di2016security} found that only 1\% of code review comments in the Chromium project were security-related. This limited presence of security feedback is largely due to factors such as the lack of security expertise among reviewers and the lower prioritization of security concerns in code reviews \cite{braz2022software}. This poses challenges for constructing high-quality datasets, as security-relevant discussions are often scattered and lack clear indicators, making them difficult to identify and extract from general review data. 

To improve the efficiency of code reviews, automated review comment generation has been extensively studied, yet most existing approaches focus on generating generic review comments rather than security-specific feedback. Code review datasets, such as those constructed by Tufano et al. \cite{tufano2022using}, primarily contain general-purpose comments and lack detailed security annotations. While some small-scale security-related review datasets exist, such as the one manually curated by Yu et al. containing 614 security-related comments \cite{yu2023security}, their limited size and diversity hinder large-scale research and model development. Furthermore, general-purpose code review models built on existing datasets may struggle to generate security-relevant feedback due to the lack of targeted training data. Therefore, automatically constructing a high-quality and scalable dataset for security-related code review becomes an urgent problem. Beyond dataset construction, evaluating the effectiveness of review models in security-related scenarios remains an open question.

To address these challenges and advance research in this area, this paper presents an approach for constructing a security-related code review dataset based on active learning. 
A review instance is considered security-related if it identifies potential security risks in the code or provides recommendations for improving code security. A fundamental challenge in this research is the absence of an initial labeled dataset, which complicates the development of an effective classifier at a low cost. To mitigate data scarcity, we employ an iterative training and annotation strategy, integrating machine-assisted processing with human feedback. Specifically, we refine the training dataset through multiple iterations, where model predictions guide the selection of data for human annotation. This process results in a training dataset comprising 3,641 annotated instances. Leveraging this dataset, we train an ensemble classifier to filter and extract security-related reviews from a raw corpus of 373,824 review instances, ultimately constructing a dataset of 6,732 security-related reviews across the programming languages C, C++, C\#, Java, and Go. With this dataset, we establish a benchmark to evaluate the performance of state-of-the-art generic code review comment generation approaches and large language models in generating security-related feedback. Experimental results indicate that while existing approaches perform well in generating general-purpose code review comments, their effectiveness in generating security-related feedback remains limited, highlighting the need for further improvement in this area. The implementation and dataset are publicly available at \cite{ReplicationPackage}.

The main contributions of this paper are as follows.
\begin{itemize}
\item We propose a framework for constructing a security-related code review dataset based on active learning. 
\item We build a high-quality, scalable dataset of security-related code reviews, consisting of 6,732 instances from multiple high-quality repositories, covering a variety of security issues and 5 programming languages. To the best of our knowledge, this is the largest publicly available dataset of security-related code reviews.
\item We evaluate the performance of state-of-the-art code review comment generation approaches in generating security-related comments, analyzing their effectiveness, and examining their limitations.
\end{itemize}

The remainder of this paper is organized as follows. Section 2 introduces the background of security vulnerabilities and active learning. Section 3 details the dataset construction approach. Section 4 presents our research question and related experimental design. Section 5 analyzes the experimental results and discusses the limitations of the study. Section 6 and 7 introduces related work and concludes the paper, respectively.
\section{Background}
\subsection{Security Vulnerabilities}
Software security vulnerabilities refer to flaws, weaknesses, or defects in a software system that can be exploited by an attacker to cause unintended or harmful behavior, potentially compromising the confidentiality, integrity, availability, or authenticity of the system or its data \cite{strout2023vulnerability}. They often arise from poor coding practices, design errors, or insufficient security measures during development. To detect security vulnerabilities early in the development process, researchers and developers have proposed various approaches, including static code analysis, dynamic code testing, and code reviews \cite{harzevili2023survey}. This paper focuses primarily on the third aspect.

To provide developers and security professionals with a reference framework, several security standards have been established, including Common Vulnerabilities and Exposures (\textbf{CVE}) \cite{vulnerabilities2005common} and Common Weakness Enumeration (\textbf{CWE}) \cite{christey2013common}. CVE is a standard for identifying and cataloging specific vulnerabilities, while CWE provides a taxonomy for identifying and classifying software vulnerabilities based on their underlying weaknesses in design, code, or implementation. Although CWE strives to remain comprehensive, it cannot cover all types of software security issues encountered in real-world scenarios. Therefore, while we refer to the CWE list as a guideline for identifying security-related review comments, we still focus on the concept of security itself. For bias analysis in our experiments, we draw inspiration from the CWE taxonomy and derive an empirical classification system through manual analysis of code review data.

\subsection{Active Learning}

\textbf{Active Learning} is an efficient machine learning approach that selects the most informative samples from a large pool of unlabeled data for annotation, leveraging data characteristics or model feedback \cite{ren2021survey, liu2022survey}. This process improves model performance while reducing the need for extensive human annotation. Compared to random sampling, active learning better optimizes the decision boundary, making it particularly effective in scenarios where annotation is costly or data distribution is highly imbalanced.

Based on data selection criteria, common active learning approaches are typically classified into two categories: \textbf{uncertainty-based} and \textbf{representativeness-based} approaches. Uncertainty-based active learning focuses on selecting samples where the model exhibits high uncertainty in its predictions. For example, it can prioritize data points with the lowest confidence scores to refine the decision boundary or select instances where multiple models generate conflicting predictions, thus leveraging the complementary insights provided by different models. In contrast, representativeness-based active learning emphasizes data diversity, ensuring that selected samples are both challenging for the model and representative of the overall data distribution, thus reducing bias and mitigating overfitting. A typical example is the coreset-based approach \cite{sener2017active}, which formulates active learning as a subset selection problem by solving a k-center clustering objective.

In this work, we incorporate ideas from both uncertainty-based and representativeness-based active learning strategies. Specifically, we utilize multiple models for decision making and select instances where the models disagree for manual annotation, ensuring that the model receives supervision in regions of highest uncertainty. Meanwhile, since positive class samples are relatively scarce, we retain all instances that are consistently classified as positive by all models to ensure sufficient coverage of the minority class. This strategy helps mitigate learning biases caused by class imbalance, thereby enhancing the model's capability in security-related code review classification tasks.
\section{Dataset Construction}
\subsection{Overview}
\begin{figure*}[htbp]
    \centering
    \includegraphics[width=0.80\textwidth]{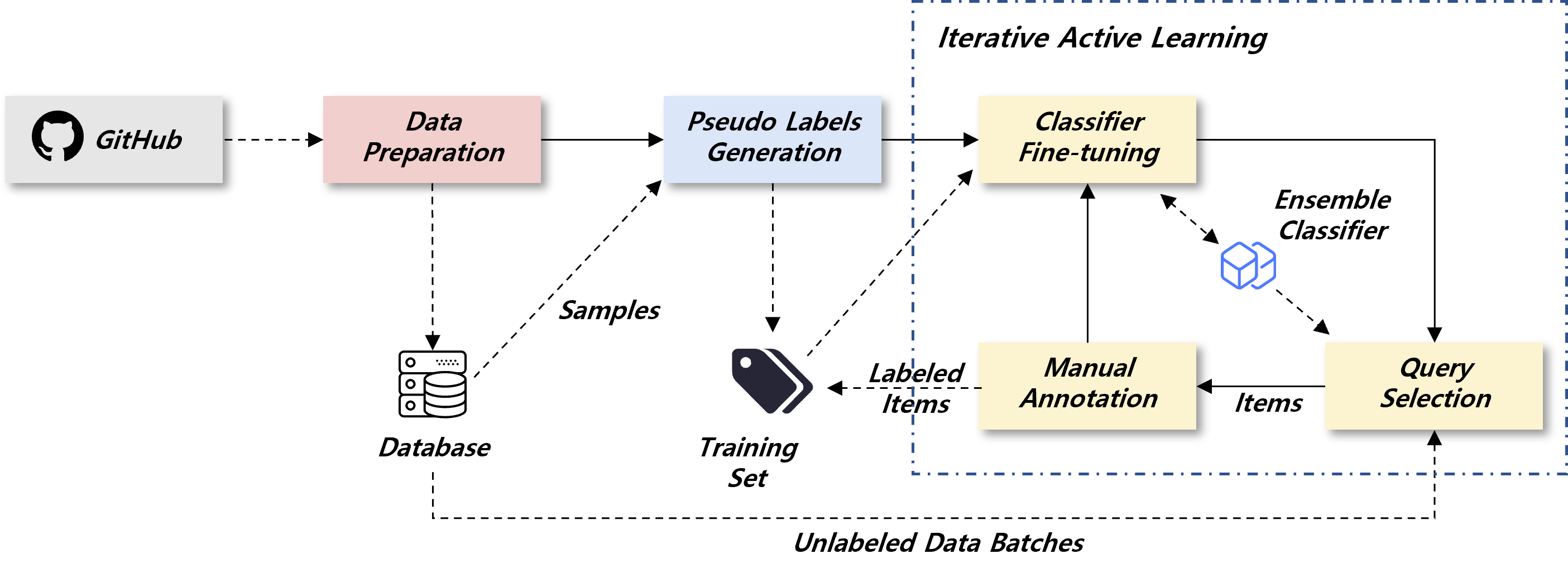}
    \caption{Overview}
    \label{fig:overview}
\end{figure*}

We proposed a dataset construction methodology (ensemble classifier) to effectively identify security-related review comments. The overview is presented in Figure~\ref{fig:overview}. The solid lines in the figure represent control dependencies, while the dashed lines represent data dependencies. It includes three parts, \textbf{Data Preparation}, \textbf{Pseudo Labels Generation}, and \textbf{Iterative Active Learning}. The input to the methodology is raw repositories crawled from code review platforms and the output is a fine-tuned ensemble classifier. The workflow is explained as follows: 
\begin{enumerate}
\item \textbf{Data Preparation}. We mine code review data from well-maintained repositories on GitHub with extensive maintenance histories to construct our initial database.
\item \textbf{Pseudo Labels Generation}. 
We randomly sample code reviews from the initial database and generate an initial \textbf{training set} through collaborative annotation between local lightweight models and remote heavyweight expert models.
\item \textbf{Iterative Active Learning}. 
The \textbf{Classifier Fine-tuning} fine-tunes the ensemble classifier using the \textbf{training set}. Next, the \textbf{Query Selection} processes data batches extracted from the database, selecting informative samples based on the predictions of the fine-tuned classifier. These selected samples are then assigned to the \textbf{Manual Annotation} for labeling, further refining the \textbf{training set}. This iterative process continues until the classifier achieves acceptable precision and recall thresholds.
\end{enumerate}

Based on the preceding steps, we develop a fine-tuned ensemble classifier. Based on this classifier, we can filter out security-related review comments from the database. 

\subsection{Data Preparation}

\subsubsection{Data Selection and Filtering Criteria}
In this study, we choose \textbf{GitHub}~\cite{GitHub} as our data source for its vast developer community, abundant code
repository resources and alignment with prior research. We mine repositories written in C, C++, C\#, Java, and Go, as these languages are more likely to involve security-related issues in practical software development. To ensure data quality, we primarily focus on the top 10,000 repositories ranked by number of stars for each language and filter out repositories with less than 1,000 pull requests. To ensure data quality and legality, we only preserve repositories with popular and distributable licenses\footnote{Apache-2.0, MIT, GPL-3.0, BSD-2.0, BSL-1.0 license, etc.}.

\subsubsection{Crawling Tools and Reviews Storage}
We utilize the \textbf{GitHub REST API} and the existing tool \textbf{ETCR}, developed by Heumüller et al. \cite{heumuller2021exploit} for crawling code review comments from GitHub. We store the crawled pull requests, commits, code review comments, and related file information in a database format to facilitate flexible querying and data extraction. Each pull request typically contains multiple commits, and each commit may affect multiple files. To optimize storage, we record the state of files before each pull request in the database and the specific changes made in each commit. This design allows us to flexibly retrieve the content and state of files before and after changes corresponding to a specific code review comment during subsequent dataset construction.

Code review often involves multi-round dialogues and the interactions between developers and reviewers during the review process generate valuable feedback. We retain these complete dialogue records, allowing data users to access them when needed. The crawling tool is designed to be scalable, allowing it to adapt to updates in GitHub repositories so the dataset we construct is also scalable.

\subsubsection{Dataset Extraction and Construction}
To facilitate dataset usage, we develop a flexible data extraction mechanism, with the extracted data format being a quintuple:
$$
\langle hunk_{diff}, file_{old}, file_{new}, comments, refinement \rangle
$$
where \textbf{$hunk_{diff}$} represents the specific code change fragment commented on by the reviewer;
\textbf{$file_{old}$} represents the old version of the file before the change;
\textbf{$file_{new}$} represents the new version of the file after the change;
\textbf{$comments$} represents the complete dialogue between developers and reviewers; and
\textbf{$refinement$} represents the optimization of the original change based on review comments, typically the file change after incorporating review feedback.

In the database, we have recorded the diff information for each commit and the base state of files at the start of each pull request. By applying each diff to the content of the base file in chronological order, we can derive the state of files at specific points in time (i.e. $file_{old}$ and $file_{new}$). We treat the most recent modification to the code fragment after the review comment in the same pull request as the refinement.

\subsection{Pseudo Labels Generation}

During the data annotation process, our primary goal was to construct an initial training dataset with clear classification boundaries to facilitate early-stage classifier training. The dataset should feature distinct positive and negative samples that are easily distinguishable, enabling the classifier to quickly learn the data features. To efficiently achieve this and reduce the manual annotation workload, we employ a model-assisted pseudo-labeling approach with a two-stage model filtering mechanism for preliminary classification.

In the first stage, we randomly extract data samples from the database and apply multiple lightweight open-source models for \textbf{zero-shot classification}. These models, typically distilled or low-parameter versions of large language models, have low computational overhead and reduced hardware resource requirements, making them cost-effective for data filtering tasks. By leveraging the diverse knowledge bases of these models, which may be trained on different datasets, we mitigate the impact of unstable factors such as model hallucinations.

In the second stage, predictions of the lightweight models are aggregated to determine positive and negative samples. A data instance is considered positive (or negative) only if all models unanimously classify it as such. These clearly classified data points and their corresponding pseudo-labels are then submitted to \textbf{expert models} for validation. Expert models, usually with larger parameters and stronger performance, provide more reliable classification results.

Real-world security-related code review instances are relatively rare. To avoid an imbalanced initial dataset and ensure that the performance of the model is not biased towards the majority class, we perform \textbf{balanced sampling} on the validated labeled samples, seeking a positive to negative ratio close to 1:1.

\subsection{Iterative Active Learning}

The pseudo-labeled dataset is used as the initial training set, and we proceed to train an ensemble classifier based on instruction-tuning. To facilitate better adaptation to the specific classification task, we utilize the local models employed during pseudo-labeling as the base models. The parameters of each model's final layer are unfrozen, and a classification head is added. The training objective is to enable the model to accurately classify code review comments as either security-related or not, based on the provided instruction and the content of the comments. 

To evaluate each model's performance, we randomly split a validation dataset from the original database and manually label it with positive and negative classes. The distribution of the validation dataset is consistent with that of real-world code review data.

The pseudo-labeled dataset is characterized by clear classification boundaries and a limited number of ambiguous instances. However, models trained on this dataset may struggle with distinguishing difficult cases, resulting that they fail to meet the desired thresholds. This is where iterative fine-tuning comes into play. Specifically, we extract unlabeled data batches from the database, pass them through each classifier, and categorize the classification results into three groups: (1) all predicted as positive, (2) inconsistent model predictions, and (3) all predicted as negative. For samples with inconsistent predictions, which are treated as difficult cases, we submit them to human annotators. In the early stages, the number of inconsistent samples is high, so we employ simple random sampling to select a subset for annotation. The sample size is determined using the following formula \cite{1977Sampling}:

\[
n = \frac{N \cdot Z^2 \cdot p \cdot (1-p)}{e^2 \cdot (N-1) + Z^2 \cdot p \cdot (1-p)}
\]
where $n$ represents the required sample size, $N$ is the total population size, $Z$ corresponds to the Z-value for the chosen confidence level, $p$ denotes the proportion of the population with a specific feature, and $e$ is the acceptable sampling error.

Given the scarcity of positive samples, it is crucial to ensure an adequate number of positive instances to maintain a balanced training dataset. Therefore, data predicted as (1) "all predicted as positive" by the model is manually confirmed and retained, while data predicted as (3) "all predicted as negative" is discarded. The manually labeled data is then incorporated into the training dataset and used for model retraining during each iteration.

\begin{figure}[t]
    \centering
    \includegraphics[width=\columnwidth]{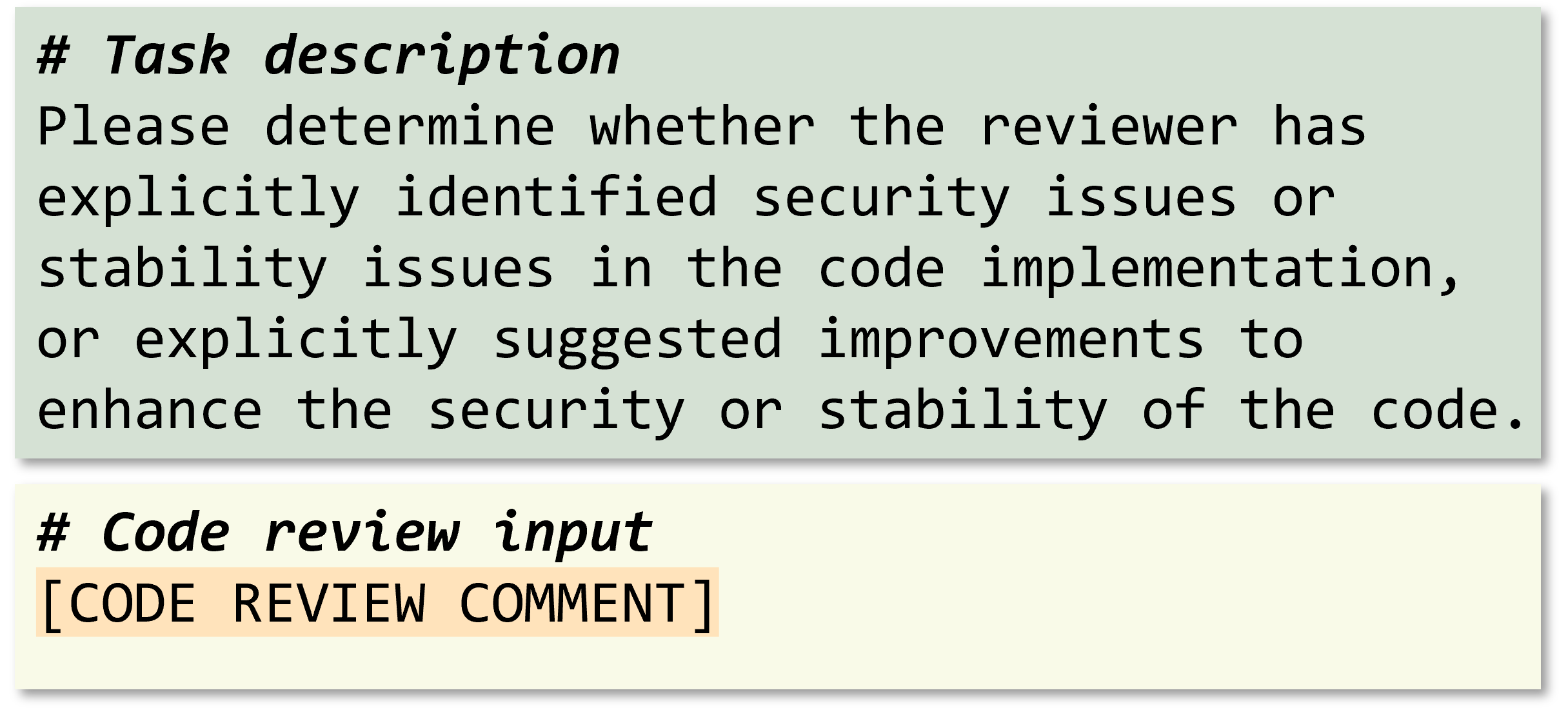}
    \caption{The prompt template used for classification}
    \label{fig:prompt}
\end{figure}

Throughout the process of \textbf{Pseudo Labels Generation} and \textbf{Iterative Active Learning}, we maintain consistent prompts to ensure that classification tasks at different stages adhere to the same input format and semantic consistency. The specific prompt content is shown in \ref{fig:prompt}. We only mention the concept of security without providing an explanation of the concept itself. This design preserves the model's understanding of security, avoiding the introduction of human biases.

\subsection{Implementation} 

\subsubsection{Data Annotation} \label{data annotation}

We invite 6 volunteers, who are proficient in software security and have experience in software review, to participate in the annotation process. The annotation process is detailed as follows: 
annotators were first familiarized with secure coding guidelines (e.g., OWASP) to establish an intensional definition of security relevance. The CWE taxonomy was used as an extensional definition—instances were labeled as security-related if they could be clearly mapped to a CWE category. In cases where no clear CWE mapping was available, labeling decisions were based on the intensional definition. Each volunteer independently assigns positive or negative labels to review items. In cases of disagreement, the volunteers discuss until reaching a consensus. We use the Fleiss' Kappa statistic \cite{kraemer1980extension} to measure the annotation agreement among different annotators. By calculating Fleiss' Kappa coefficient (0.88), we observe a high level of agreement among annotators, indicating that the accuracy and reliability of the manual annotations are acceptable. 

\subsubsection{Data Split and Classifier Training}

We randomly divide our entire unlabeled dataset from the database into 100 batches, with the batch size set to 3,776. We use the first 15 batches for iterative training and randomly preprocess another 2 batches as the validation set and test set, with sizes of 3,743 and 3,727, respectively. Following traditional training principle, the validation dataset is used for model training and test dataset is used to assess the model's generalization ability. Due to considerations of context size limitations, classifier performance, and annotation cost, we focus solely on the first comment from the reviewer in dialog-type comments for classification. This applies to model training, inference and human annotation. We employ a diverse set of advanced models: Gemma-2-9B-IT \cite{gemma_2024}, Meta-Llama-3-8B-Instruct \cite{llama3modelcard}, DeepSeek-7B-Chat \cite{deepseek-llm}, Ministral-8B-Instruct-2410 \cite{ministral8b}, and Qwen2.5-7B-Instruct \cite{qwen2.5} as base models. They are chosen for their strong performance and representativeness among open-source models. We utilize \textbf{DeepSeek-V3} \cite{deepseekai2024deepseekv3technicalreport} as the expert model and construct an initial pseudo-labeled training dataset of 3,000 samples, maintaining a balanced ratio of 1:1 between positive and negative samples. The epochs and learning rate during the training stage are set to 30 and 0.00005. We use AdamW optimizer with linear warmup of 500 warmup steps. In the collaborative iteration process, a total of 3,641 training samples are annotated. As high precision is required during training, the evaluation function is adapted to a dynamic weighted F1 score, defined as follows:
\[
     \text{F1}_{\text{dynamic}} = \frac{(1 + \alpha) \cdot (w_p \cdot \text{precision}) \cdot (w_r \cdot \text{recall})}{w_p \cdot \text{precision} + \alpha \cdot w_r \cdot \text{recall}}
\]

where \( \alpha \) is a hyperparameter set to 2 in our experiments, and \( w_p \) and \( w_r \) are threshold penalty parameters designed to penalize lower precision and recall values. For the parameters in the sampling formula, we set Z-value to 1.96 corresponding to 95\% confidence, p to 0.5 to maximize the sample size and e to 0.05.

\subsubsection{Query Strategy}

To determine the most effective query strategy during inference, we evaluate multiple voting schemes on validation set and compare their impact on model performance. Based on these experimental results, we select the 4-out-of-5 strategy as the optimal choice—labeling a sample as positive only if at least 4 out of 5 models agree; otherwise, it is marked as negative. Notably, this strategy is used consistently in classifier evaluation and dataset construction.
\section{Study Design}
\subsection{Research Questions}
To evaluate the quality of the proposed dataset and the capability of existing code review comment generation methods in generating security-related review comments, we investigate the following research questions:

\noindent\textbf{RQ1: What percentage of code review comments are security-related, and how are they categorized?} 

\noindent RQ1 serves as the foundation of our study. Due to the scarcity of security-related code review comments, manually constructing such a dataset is challenging. In order to better analyze the composition of the constructed dataset, we categorize security-related code review comments in real-world scenarios.

\noindent\textbf{RQ2: How accurate is the proposed approach in identifying security-related code review comments, and how does it compare to existing approaches?} 

\noindent Since the construction of the dataset relies on the classification approach, we aim to evaluate its precision and recall to ensure the reliability of the dataset.

\noindent\textbf{RQ3: How does Sere dataset compare to real-world scenarios? Does it preserve high precision and introduce low bias?} 

\noindent We aim to validate the quality of the dataset through a sampling-based evaluation, focusing primarily on metrics such as precision and representativeness.

\noindent\textbf{RQ4: How do state-of-the-art code review comment generation approaches perform on Sere dataset?}

\noindent RQ4 evaluates the state of the art in automated generation of security-related code reviews with our large-scale dataset. The evaluation results motivate further research in automated generation of security-related code reviews.

\subsection{Empirical Category System}

Based on observations and summarization of positive instances in the test dataset, we construct a classification system of 14 categories. Unlike prior work relying on predefined categories to collect data, we derived categories from actual data by mapping review content to CWE labels or secure coding keywords, then semantically grouping them. Categories were named for conciseness and broad coverage to support downstream evaluation. The security-related data are submitted to human annotators and assigned a corresponding category label. In cases of disagreement,
the volunteers discuss until reaching a consensus. The calculated kappa coefficient reaches 0.81. We refer to this labeled subset as the \textbf{representative set}, which is used for \textbf{RQ1} and \textbf{RQ3}.

\subsection{Classifier Evaluation}

We first evaluated the recall of our ensemble classifier on the manually constructed datasets by Yu et al.\cite{yu2023security} and Paul et al.\cite{paul2021security}, respectively.
Then we compare the following approaches with our classifier: keyword matching based on 105 collected security-related keywords, a traditional technique commonly used in related work \cite{paul2021security, paul2022astor, yu2023security}; LLMs including \textbf{GPT-4o} \cite{GPT-4o}, \textbf{Gemini-1.5-Pro} \cite{team2024gemini}, and \textbf{DeepSeek-V3}, using the prompt shown in Figure \ref{fig:prompt}. We employ accuracy, precision, recall, and F1 score as evaluation metrics. These metrics reflect the overall correctness of the ensemble classifier, its precision in identifying security-related comments, and its coverage of actual security-related comments, respectively.

\subsection{Sampling-based Dataset Analysis} 
We first randomly sample 447 items (guaranteeing a confidence of 95\%) from our dataset and perform human annotation following the process in Section \ref{data annotation}. The annotators independently annotate each sample as security-related or not and resolve disagreements through discussion. We then calculate the precision of the dataset based on the annotation results. To evaluate potential bias in the constructed dataset, we choose Fisher's Exact Test \cite{fisher1970statistical} because it provides reliable evaluation results when dealing with small datasets, well-suited for sparse security-related code review data scenarios. We conduct tests based on sampling. Each time, we randomly sample a set of the same size as the representative set from the constructed dataset. The samples are then manually annotated with security categories following the annotation process of representative set, and Fisher’s Exact Test is performed between the labeled sample data and the representative set. We repeat this process three times, recording the p-value for each subcategory in each test. We set the significance level \(\alpha = 0.05\), which means that we accept a 5\% probability of rejecting the null hypothesis (there is no distribution difference between the sampled data and the representative set) when it is actually true. 

\subsection{Generation of Security-related Code Reviews} 

\subsubsection{Subject Selection}
In RQ4 evaluation phase, we focus on the performance of SOTA open-source code review comment generation models as baselines: \textbf{Auger}\cite{li2022auger}, \textbf{DISCOREV} \cite{ben2024improving}, \textbf{LLaMA-Reviewer} \cite{lu2023llama} and \textbf{CodeReviewer} \cite{li2022automating}. For CodeReviewer, LLaMA-Reviewer and Auger, the original authors provide pre-trained model checkpoints, which we directly use for experimentation. Specifically, the LLaMA-Reviewer model includes two fine-tuning versions: prefix-tuning and LoRA-tuning. For DISCOREV, the authors provide complete reproducible package, allowing us to replicate the approach based on the original code and training settings and evaluate the reproduced model. Since AUGER is designed for Java, to ensure a fair comparison, we select Java language scenarios in Sere as the evaluation dataset for Auger. Large language models (LLMs) have demonstrated strong capabilities in various software engineering tasks, including code generation and review. However, their effectiveness in generating security-related code review comments remains unclear. Therefore, we select \textbf{GPT-4o} and \textbf{DeepSeek-V3} as representative LLMs and evaluate their performance under both zero-shot and few-shot settings.

\subsubsection{Data Preprocessing}

To better evaluate these code review comment generation approaches, we preprocess the dataset following the methodology proposed by Tufano et al.~\cite{tufano2022using}, including removing non-English data, replacing links in comments, and discarding excessively long code changes or contexts. Notably, samples where the number of code change tokens exceeds 2,048 or context tokens exceeds 10,000 are removed. As a result, we curate a benchmark of 4,788 evaluation samples from the original 6,731 data points. Since the objective of this evaluation is to assess the single-turn generation performance of these approaches, we retain only the first comment from the reviewer in the dialogue data.

\subsubsection{Prompt Design}
\begin{figure}[t]
\centering
\includegraphics[width=\columnwidth]{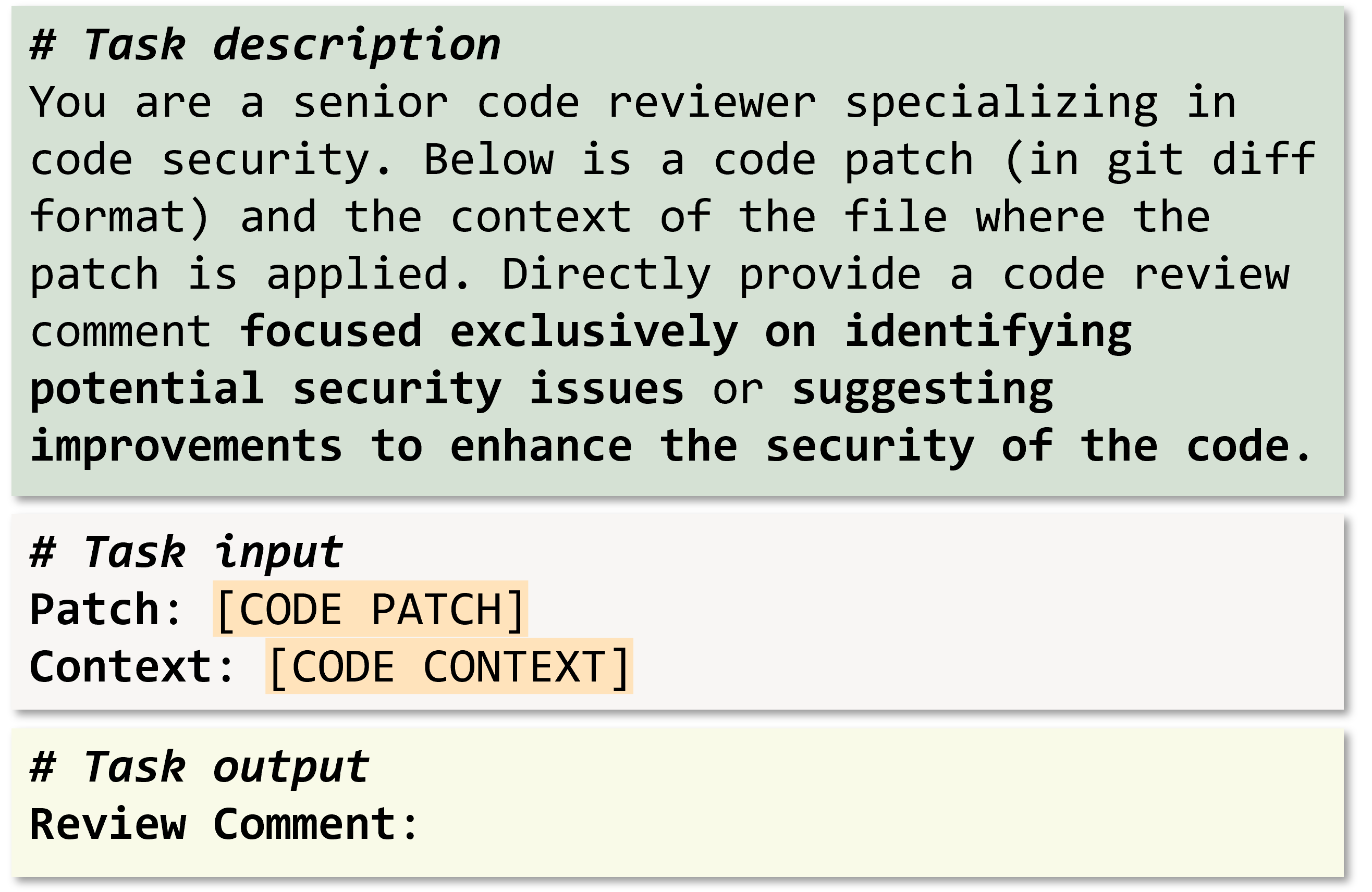}
\caption{The prompt used for zero-shot comment generation}
\label{fig:prompt_zero_shot}
\end{figure}

\begin{figure}[t]
    \centering
    \includegraphics[width=\columnwidth]{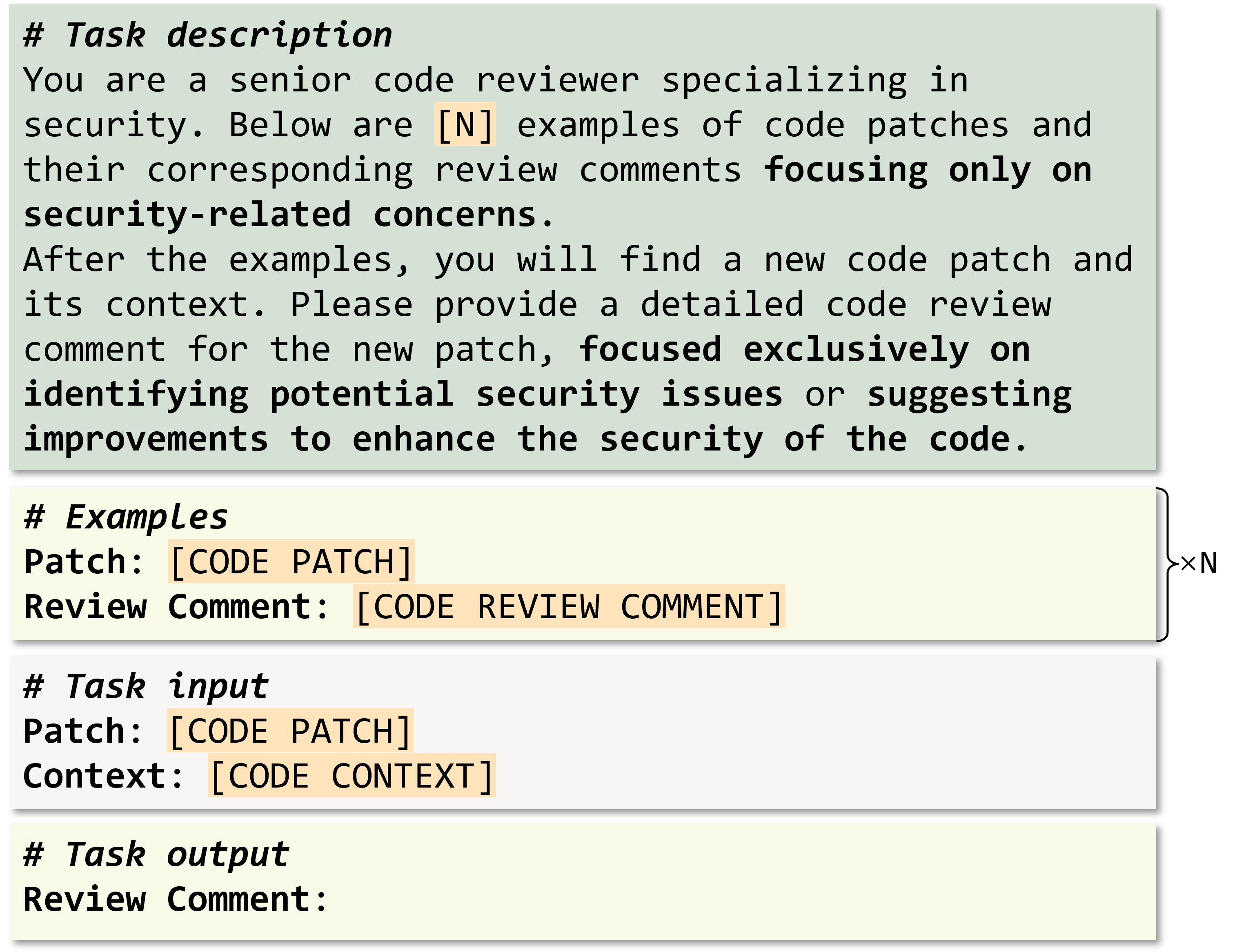}
    \caption{The prompt used for few-shot comment generation}
    \label{fig:prompt_few_shot}
\end{figure}

We design prompt strategies for LLMs under both zero-shot and few-shot settings. In the actual few-shot evaluation, we select two samples from repositories distinct from the one containing the original data as few-shot examples. Due to the extensive length of contextual information, we opt not to include it in the few-shot examples to avoid potential dilution of task-relevant features and ensure a more focused learning signal for the model. The prompt templates we use are shown in Figure \ref{fig:prompt_zero_shot} and \ref{fig:prompt_few_shot}.

\subsubsection{Metrics} 

The performance of the review generation approaches is assessed using several metrics, including \textbf{BLEU} \cite{papineni2002bleu}, \textbf{ROUGE-L} \cite{lin2004rouge}, \textbf{METEOR} \cite{banerjee2005meteor}, \textbf{Exact Match}, and \textbf{Semantic Similarity}. BLEU, ROUGE-L, METEOR, and Exact Match are standard evaluation metrics that measure different aspects of similarity between the generated comments and the ground truth. BLEU and ROUGE-L primarily capture lexical and structural similarities, METEOR incorporates semantic matching through stemming, synonymy, and paraphrase recognition, while Exact Match evaluates whether the generated comment is identical to the reference comment. For Semantic Similarity, we use \textbf{Sentence-BERT (SBERT)} \cite{reimers-2019-sentence-bert} to generate embeddings for both the generated and reference comments, then compute the \textbf{cosine similarity} between these embeddings to measure their semantic similarity. This approach allows us to capture the semantic closeness between the generated and reference comments, overcoming the limitations of traditional token-based metrics by considering the meaning rather than exact word matches.

\subsection{Experimental Environment} 

We conduct our experiments on Ubuntu 22.04.5 LTS operating system, with the training framework implemented using PyTorch. The hardware configuration consists of two NVIDIA A800 80GB GPUs to support model training and inference. For all operations requiring access to LLMs, we utilize the official APIs.
\section{Results Analysis}

\subsection{RQ1: Proportion and Categorization of Security-Related Comments}
In RQ1, we investigate the proportion of security-related comments within code review feedback and their internal composition. Through an in-depth analysis of a representative dataset comprising 3,727 code review comments, we identified 151 security-related instances, accounting for 4.05\% of the total comments.

We conducted a manual analysis of the security-related data and developed a empirical classification system, categorizing the security-related comments into 14 categories. These categories include memory management, concurrency and synchronization, exception handling, resource management (referring to the management of non-memory resources such as file handles or network connections), input validation, sensitive information management, dangerous API usage, undefined behavior (where the code's behavior is not well-defined by the language or platform specifications), permission and authentication, formatting and string issues, data protection, arithmetic issues, type confusion, and others (which include comments that do not fit into the predefined subcategories or where the security concern is too vague to be classified under a specific label). 

\begin{figure}[t]
    \centering
    \includegraphics[width=\columnwidth]{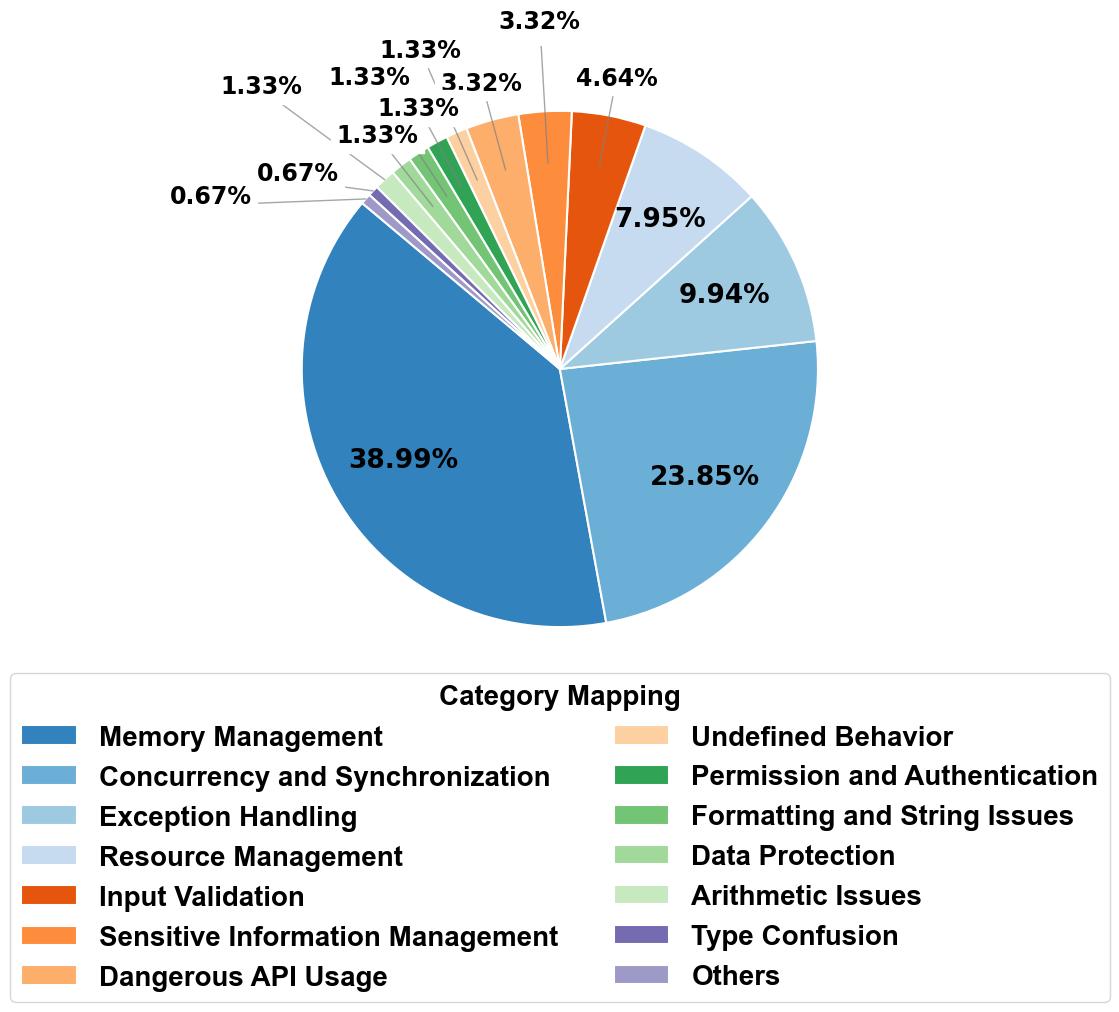}
    \caption{Distribution of security-related code review comments across 14 subcategories.}
    \label{fig:distribution}
\end{figure}

Figure~\ref{fig:distribution} illustrates the distribution of these categories within the security-related comments. The results reveal that \textbf{memory management} (38.99\%) and \textbf{concurrency and synchronization} (23.85\%) are the most prevalent categories, collectively accounting for over 60\% of the security-related comments. Other notable categories include exception handling (9.94\%) and resource management (7.95\%). In contrast, categories such as arithmetic issues, type confusion, and others have minimal representation, each constituting less than 2\%. This distribution highlights common security issue patterns identified during code reviews, particularly the frequent attention given to memory management and concurrency issues, which are often overlooked during development but receive significant focus during code review.

\begin{findingbox}{Answer to RQ1}
Security-related code review comments account for approximately \textbf{4\%} of the total comments, with the majority primarily addressing memory management and concurrency synchronization issues.
\end{findingbox}

\subsection{RQ2: Performance of Dataset Construction Approach}

\begin{table}[t]
\centering
\caption{Performance Comparison of Different Approaches}
\begin{tabular}{lcccc}
\hline
\textbf{Approach} & \textbf{Precision} & \textbf{Recall} & \textbf{Accuracy} & \textbf{F1} \\
\hline
Keyword & 0.1357 & 0.1258 & 0.9321 & 0.1306 \\
Gemini-1.5-Pro & 0.5482 & 0.7152 & 0.9646 & \textbf{0.6207} \\
GPT-4o & 0.4158 & \textbf{0.7682} & 0.9469 & 0.5395 \\
DeepSeek-V3 & 0.5968 & 0.4901 & 0.9659 & 0.5382 \\
\textbf{Ensemble} & \textbf{0.9275} & 0.4238 & \textbf{0.9753} & 0.5818 \\
\hline
\end{tabular}
\label{tab:classifier_performance}
\end{table}

The recall rates on datasets proposed by Yu et al.\cite{yu2023security} and Paul et al.\cite{paul2021security} are 75\% and 85\%, respectively. As shown in Table~\ref{tab:classifier_performance}, the experimental results demonstrate that our ensemble approach achieves the highest precision (92.75\%) and accuracy (97.53\%), significantly outperforming both traditional (keyword matching) and LLM-based approaches. Notably, high precision is crucial for ensuring the reliability and quality of the dataset, as it minimizes false positives.
In contrast, GPT-4o and Gemini-1.5-Pro exhibit higher recall rates (76.82\% and 71.52\%, respectively), but their precision is considerably lower (41.58\% and 54.82\%), indicating a higher rate of false positives in identifying security-related comments. The keyword matching approach performs the worst, with both precision (13.57\%) and recall (12.58\%) significantly lower than the other approaches.

These results validate the advantages of our ensemble approach in terms of precision and accuracy, effectively reducing false positives and ensuring the quality of the constructed dataset. The relatively lower recall reflects a deliberate design choice to prioritize precision, which may result in some security-related comments being missed. This trade-off is intentional, as our primary goal is to build a high-quality security-related dataset rather than maximizing the classifier's recall. Compared to existing approaches, our approach demonstrates superior precision, validating the effectiveness of ensemble learning and human annotation in enhancing classification performance. Additionally, the higher recall of GPT-4o and Gemini-1.5-Pro suggests that LLMs have potential in identifying potential security-related comments, although their high false positive rates limit their reliability in practical applications.

To further investigate the trade-off between recall and precision, we sample security-related review comments missed by the optimal voting strategy using alternative strategies. We then analyze the characteristics of these false negatives. The most frequent types of missed cases include: \textbf{Form Interleaving} (where the security context is obscured due to interleaving of code and natural language); \textbf{Topic Interleaving} (where security concerns are blended with non-security issues, making them less prominent); and \textbf{Rare Categories} (underrepresented types of security issues that often depend on uncommon, repository-specific contexts).

\begin{findingbox}{Answer to RQ2}
Our approach achieves high precision while maintaining acceptable recall,  ensuring the quality and robustness of the constructed dataset.
\end{findingbox}
\subsection{RQ3: Dataset's Quality Analysis}

We conduct our analysis from 4 perspectives. We first validate the precision of our dataset and present a concise comparison between our dataset and existing security-related dataset. Based on the categories in \textbf{RQ1}, we analyze the composition and potential biases of the dataset.

The human evaluation results suggest that 94\% of the samples have been correctly labeled, which indicate that the resulting dataset is relatively reliable. If a code review maps to a CWE category, we treat it as a real vulnerability. The sampling-based validation showed that 84\% of security-related reviews correspond to real vulnerabilities. The rest include insecure practices or unmapped vulnerabilities. We then compare \textbf{Sere} with existing security-related code review datasets in terms of dataset size, diversity, bias, and scalability. Sere contains 6,732 review items, significantly more than the sizes of Yu’s dataset (614) and Paul’s dataset (516). Sere is sourced from 311 high-star, multi-language projects on GitHub, whereas Yu’s dataset is derived from 4 projects within the OpenStack and Qt communities, and Paul’s dataset is sourced from the Chromium OS project. Yu’s and Paul’s datasets are subject to biases introduced by predefined security-related keywords. In contrast, Sere is constructed based on a principled definition of security, without any preset preferences towards specific types of security issues. Yu’s and Paul’s datasets rely heavily on manual annotation and post-hoc human review, whereas the construction of Sere is more automated, resulting in greater scalability.

Our dataset is derived from open-source projects and encompasses a variety of programming languages, including C, Go, C\#, Java, and C++, which are widely representative in the open-source community. Table~\ref{tab:data_distribution} presents the distribution of repositories, code review entries, and individual comments (a single code review entry may contain multiple comments) across different programming languages in both the raw dataset and our constructed dataset. The distribution of repositories, reviews, and comments across these languages remains relatively balanced, with no single language overwhelmingly dominating the dataset. The relative proportions of different languages in the constructed dataset generally align with those in the raw dataset. This suggests that our filtering criteria does not disproportionately favor or excludes specific languages, preserving the diversity and representativeness of the original dataset while ensuring a focused subset for security-related code review data.

\begin{table}[t]
\centering
\caption{Review Data Distribution Across Programming Languages: From Raw Data to Constructed Dataset} 

\begin{tabular}{l|c|c|c|c|c|c}
\hline
\multirow{2}{*}{\textbf{Lang.}} & \multicolumn{2}{c|}{\textbf{\#Repos.}} & \multicolumn{2}{c|}{\textbf{\#Reviews}} & \multicolumn{2}{c}{\textbf{\#Comments}} \\
\cline{2-7}
& Raw & Filtered & Raw & Filtered & Raw & Filtered \\
\hline
C & 62 & 55 & 37,555 & 1,534 & 74,120 & 3,312 \\
C\# & 90 & 75 & 103,545 & 1,454 & 198,929 & 3,436 \\
C++ & 53 & 51 & 81,910 & 1,269 & 154,516 & 2,824 \\
Go & 103 & 88 & 92,471 & 1,358 & 177,429 & 3,087 \\
Java & 51 & 42 & 58,343 & 1,117 & 107,114 & 2,473 \\
\hline
\textbf{Total} & 359 & 311 & 373,824 & 6,732 & 712,108 & 15,132 \\
\hline
\end{tabular}
\label{tab:data_distribution}
\end{table}

To verify whether the distribution of the constructed dataset aligns with real-world scenarios, we conduct multiple samplings from the constructed dataset, denoted as S.1, S.2 and S.3 in Table~\ref{tab:component_analysis}, and assess the alignment between the samples and the representative set in the category distribution using Fisher's exact test. The results reveal that for the categories related to \textbf{concurrency and synchronization} and \textbf{exception handling}, the p-values are notably small, suggesting that there might be subtle distribution shift for these categories. However, despite these potential variations, the tests pass successfully, and the null hypothesis is not rejected, indicating no clear evidence of significant differences in the distributions. For most other categories, the p-values are significantly higher than the predetermined significance level of 0.05. This reinforces the reliability of the data and the robustness of the null hypothesis, further suggesting that the constructed dataset's distribution aligns well with real-world security-related code review data.

\begin{findingbox}{Answer to RQ3}
The constructed dataset demonstrates negligible bias and closely aligns with real-world data distributions, ensuring its representativeness and practical applicability.
\end{findingbox}
\subsection{RQ4: Performance of Review Generation Approaches}

\begin{table}[t]
\centering
\caption{Component Bias Analysis Results}
\begin{tabular}{l|p{0.1\linewidth}|p{0.1\linewidth}|p{0.1\linewidth}}
\hline
\multicolumn{1}{c|}{\multirow{2}{*}{\textbf{Category}}} & \multicolumn{3}{c}{\textbf{Test Result (p-value)}} \\
\cline{2-4}
& S.1 & S.2 & S.3 \\
\hline
Formatting and String Issues & 1.0 & 1.0 & 1.0 \\
Type Confusion & 1.0 & 1.0 & 1.0 \\
Permission and Authentication & 1.0 & 1.0 & 1.0 \\
Undefined Behavior & 1.0 & 1.0 & 0.68 \\
Input Validation & 1.0 & 0.62 & 0.77 \\
Memory Management & 0.73 & 0.55 & 0.48 \\
Concurrency and Synchronization & 0.51 & 0.06 & 0.12 \\
Resource Management & 1.0 & 0.34 & 0.65 \\
Dangerous API Usage & 0.21 & 0.72 & 0.45 \\
Sensitive Information Management & 1.0 & 1.0 & 0.21 \\
Arithmetic Issues & 0.68 & 0.1 & 0.45 \\
Exception Handling & 0.54 & 0.12 & 0.07 \\
Data Protection & 1.0 & 0.5 & 1.0 \\
Others & 1.0 & 1.0 & 1.0 \\
\hline
\end{tabular}
\label{tab:component_analysis}
\end{table}

\begin{table*}[t]
\centering
\caption{ Performance of Code Review Generation Approaches}

\begin{tabular}{llllll}
\hline
\textbf{Approach} & \textbf{BLEU} & \textbf{ROUGE-L} & \textbf{METEOR} & \textbf{Exact Match} & \textbf{Semantic Similarity}\\
\hline
Auger & 3.67 (15.79) & 11.72\% (22.04\%) & 6.25\% (18.59\%) & 0.0\% (4.14\%) & 16.36\% (28.43\%) \\

DISCOREV & 3.22 (7.42) & 9.37\% (13.89\%) & 6.9\% (13.51\%) & 0.00\% (0.24\%) & 16.70\% (29.81\%) \\
LLaMA-Reviewer\textsuperscript{\scalebox{0.7}{Prefix}} & 3.67 (5.19) & 10.25\% (11.58\%) & 8.62\% (10.83\%) & 0.00\% (0.0\%) & 23.50\% (24.45\%) \\
LLaMA-Reviewer\textsuperscript{\scalebox{0.7}{Lora}} & 3.12 (5.7) & 7.82\% (10.39\%) & 5.38\% (8.29\%) & 0.00\% (0.16\%) & 17.24\% (22.87\%) \\
CodeReviewer & 3.15 (5.44) & 7.73\% (10.18\%) & 5.53\% (8.29\%) & 0.02\% (0.06\%) & 22.09\% (23.17\%) \\
DeepSeek-V3\textsuperscript{\scalebox{0.7}{0-shot}} & 0.98 & 5.87\% & 13.7\% & 0.00\% & 39.82\% \\
DeepSeek-V3\textsuperscript{\scalebox{0.7}{2-shot}} & 0.81 & 5.03\% & 12.33\% & 0.00\% & 39.43\% \\
GPT-4o\textsuperscript{\scalebox{0.7}{0-shot}} & 0.89 & 5.37\% & 13.43\% & 0.00\% & 38.76\% \\
GPT-4o\textsuperscript{\scalebox{0.7}{2-shot}} & 0.66 & 4.42\% & 11.78\% & 0.00\% & 38.18\% \\
\hline
\end{tabular}
\label{tab:generation_performance}
\end{table*}

The performance of code review comment generation approaches on our benchmark is summarized in Table~\ref{tab:generation_performance}. For conciseness, we refer to the first 5 models as the "generic code review models" and generation based on DeepSeek and GPT as the "large language model (LLM)-based approach". For generic code review models, their performance on the test dataset used in the original papers is reported in parentheses for comparison. In a multi-programming-language scenario, the prefix version of LlamaReviewer achieves the highest BLEU score (3.67) and the highest ROUGE-L score (10.25\%), while DeepSeek-V3 based on zero-shot prompt performs best in terms of METEOR(13.7\%) and Semantic Similarity (39.82\%). However, all approaches perform poorly on Exact Match, with results nearly approaching zero, indicating that they struggle to generate comments that exactly match the ground truth. 

In general, generic code review models perform well in lexical similarity, such as BLEU and ROUGE-L. This may be attributed to the fact that these approaches primarily rely on such metrics during training as evaluation criteria, leading to generated comments that conform well to real-world code review formats. In contrast, LLMs achieve better results on semantic-based metrics such as METEOR and Semantic Similarity but fall short in lexical similarity. This could be due to the lack of fine-tuning, leading to more varied word choices while still preserving accurate semantic expression. In particular, the zero-shot and few-shot settings for LLMs show minimal differences in performance, with zero-shot settings slightly outperforming few-shot settings in most cases. This indicates that providing additional examples may not necessarily improve the performance of LLMs in this task.

When applied to the security-related evaluation dataset,  generic code review models experience a noticeable performance drop, compared to their publicly available experimental results, indicating their limitations in handling security-specific contexts. This is largely due to the scarcity of security-related data, which constitutes only a small portion of the training set, preventing the models from effectively handling this task. Furthermore, these approaches are constrained by their parameter capacity and knowledge scope: without security-related data for fine-tuning, they fail to transfer relevant knowledge to this task. In contrast, large language models, benefiting from extensive training corpora, demonstrate strong transferability. However, there remains a considerable gap between the comments generated by LLMs and those written by human reviewers.

\begin{figure}[t]
    \centering
    \includegraphics[width=\columnwidth]{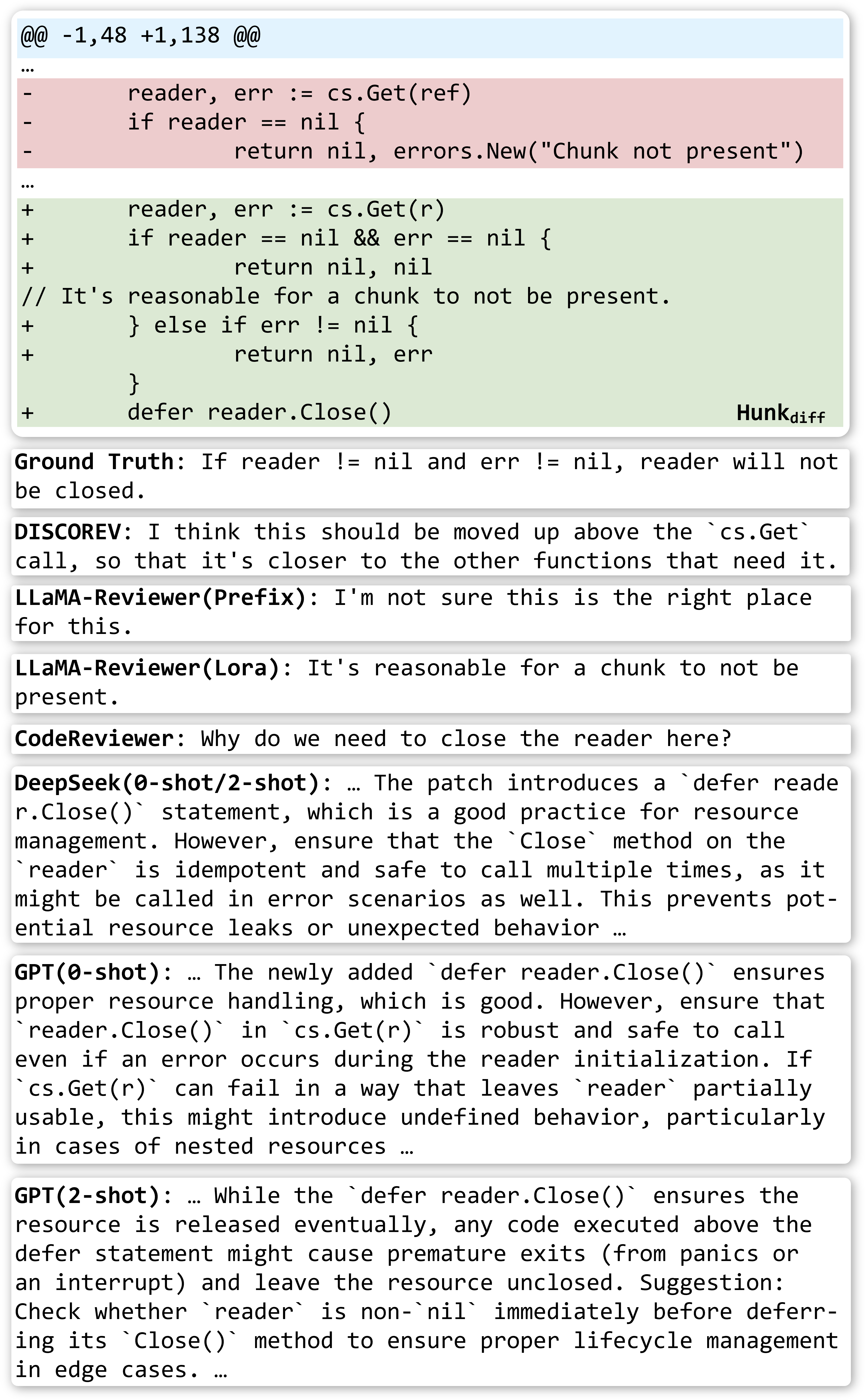}
    \caption{Illustrative Example.}
    \label{fig:example}
\end{figure}

\textbf{Illustrative Example.} To provide a clear and intuitive understanding of the effectiveness of different approaches in security-related code review comment generation, we present a simple example shown in Figure \ref{fig:example}. Due to space limitations, we focus on the $hunk_{diff}$ part in the task input. Since large language models often produce lengthy responses, we extract and display the parts of the results that are most relevant to the ground truth. Additionally, since the responses from DeepSeek (0-shot and 2-shot) are quite similar, we present only one of them.

The original code presents a potential security issue in the form of improper resource deallocation. The ground truth highlights that if both reader != nil and err != nil, reader should not be closed. DISCOREV focuses on reorganizing the function calls but misses the core security issue. LLaMA-Reviewer (Prefix) expresses uncertainty about the placement of the defer statement, while LLaMA-Reviewer (Lora) overlooks the security concern entirely, directly copying the code comment. CodeReviewer raises a question about the necessity of closing the reader but does not address the specific security vulnerability. DeepSeek-V3 (0-shot/2-shot) acknowledges the importance of resource management, noting the need for idempotent "Close" calls, but fails to recognize the specific condition where the resource may not be closed properly due to premature exits. GPT-4o (0-shot) focuses on undefined behavior rather than addressing the resource deallocation issue under specific conditions. Only GPT-4o (2-shot) offers a solution that directly addresses the potential risk of premature exits, suggesting checking reader before deferring the Close() approach to ensure proper resource management.

\begin{findingbox}{Answer to RQ4}
LLMs achieve higher accuracy in semantic expression, while generic code review generation approaches maintain more precise word choices, suggesting room for improvement in both approaches for generating effective security-related feedback.
\end{findingbox}

\subsection{Threats to Validity}
\textbf{Internal Validity.}
The first threat is data leakage, where the evaluation benchmark may contain data that overlaps with the training sets of the evaluated approaches. To mitigate this, we preprocess the data, which may differ significantly in form from the original training data following ~\cite{tufano2022using}. Furthermore, Cao et al. \cite{cao2024concerned} finds that contaminated data does not always impact model performance, and there are instances where models perform even better when evaluated on data collected after their cutoff date. Another threat is the design of prompts for evaluating large language models. To minimize bias, we design prompts to be as clear and direct as possible, avoiding ambiguous or misleading expressions that could influence the models' outputs. Finally, the reproducibility of classifier training is a concern due to the randomness in sampling during the dataset construction process. However, we have retained the final constructed training dataset, ensuring that others can reproduce and validate our results.

\noindent\textbf{External Validity.}
The first threat is the selection of models in our approach and evaluation. To address this, we choose effective models that are commonly used in the research community.
Another potential threat comes from data sources, as different sources may yield varying results. To mitigate this, we select widely acknowledged repositories as our sources. Finally, the reproducibility of the LLMs’ outputs is a concern, as their behavior can vary due to randomness or other factors in the generation process. To mitigate this, we retain the generated outputs from the LLMs, ensuring that our results could be independently verified.

\noindent\textbf{Construct Validity.}
The first threat is the potential influence of personal bias in human annotation. To address this, we measure inter-annotator agreement using the kappa statistic to ensure consistency.
Another threat is the limitation of code review generation evaluation metrics. We choose commonly used metrics in code review comment generation tasks. In future work, we plan to incorporate additional metrics to provide a more comprehensive assessment of model performance.

\section{Related Work}
\subsection{Automated Code Review Generation}
The automation of code review generation has evolved from retrieval-based techniques to advanced deep learning and large language models (LLMs) \cite{chen2025deep}. Early approaches, such as DeepMem by Gupta et al. \cite{gupta2018intelligent}, employed LSTM models to retrieve relevant historical reviews, while Siow et al. \cite{siow2020core} enhanced semantic understanding using attention mechanisms. Hong et al. \cite{hong2022commentfinder} introduced Gestalt Pattern Matching in CommentFinder to identify similar approaches via cosine similarity. Despite their contributions, these approaches were constrained by reliance on historical data and the absence of generative capabilities.

The introduction of deep learning marked a shift toward generative models. Tufano et al. \cite{tufano2022using} leveraged the Text-To-Text Transfer Transformer (T5) to generate review comments directly from source code, demonstrating the efficacy of pre-trained models in this task. Li et al. \cite{li2022automating} developed CodeReviewer, incorporating four pre-training tasks specific to code review, leading to improvements in review comment generation and code change quality estimation. AUGER \cite{li2022auger} further enhanced precision by explicitly linking generated comments to specific code lines.

Recent advancements in LLMs have further improved automated code review. Lu et al. \cite{lu2023llama} introduced LLaMA-Reviewer, utilizing Parameter-Efficient Fine-Tuning (PEFT) to achieve competitive performance with fewer trainable parameters. Nashaat et al. \cite{nashaat2024towards} proposed CodeMentor, incorporating reinforcement learning with human feedback (RLHF) to align model outputs with domain-specific requirements. Yu et al. \cite{yu2024fine} introduced Carllm, emphasizing accuracy and clarity in review comments by providing precise locations, explanations, and repair suggestions. Ben et al. \cite{ben2024improving} proposed DISCOREV, employing cross-task knowledge distillation to integrate review tasks such as quality estimation, comment generation, and code refinement, achieving state-of-the-art BLEU and CodeBLEU scores. These advancements underscore the increasing role of LLMs and domain-specific adaptations in improving automated code review quality and relevance.

\subsection{Security-Related Code Review}
The role of code review in software security has been widely studied, highlighting its strengths and limitations in vulnerability detection. Di Biase et al. \cite{di2016security} found that only 1\% of Chromium review comments addressed security, often overlooking language-specific issues like C++ memory errors and XSS. Paul et al. \cite{paul2021security} identified factors influencing security defect detection in Chromium OS, such as review time, reviewer experience, and code complexity, stressing the risks of tangled code changes and reviewer fatigue.

Thompson and Wagner \cite{thompson2017large} analyzed 3,126 open-source projects, finding a strong correlation between thorough reviews and reduced vulnerabilities, though the number of reviewers had little impact. Yu et al. \cite{yu2023security} showed that race conditions and resource leaks were frequently flagged, while XSS and SQL injection were often missed, emphasizing the need for clearer reviewer guidance.

To improve security-focused reviews, Paul \cite{paul2022astor} developed a classifier for identifying security-related reviews, while Charoenwet et al. \cite{charoenwet2024empirical} found that combining SAST tools increased vulnerability detection but produced 76\% irrelevant warnings. Yu et al. \cite{yu2024insight} evaluated LLMs in security review, showing that models like GPT-4 outperform static analysis tools but struggle with verbosity and logical inconsistencies, with prompt design significantly affecting performance.

\subsection{Code Review Dataset}

Existing research has introduced several datasets for studying code review, which can be broadly categorized into general-purpose and security-related datasets. For general-purpose datasets, Tufano et al. \cite{tufano2022using} compiled a dataset integrating Stack Overflow discussions and the CodeSearchNet Java dataset, producing over 1.48 million instances after preprocessing. Li et al. \cite{li2022auger} constructed another dataset from 11 Java repositories on GitHub, selecting projects with at least 4,000 pull requests and 100 contributors. Extending beyond Java, Li et al. \cite{li2022automating} introduced a multilingual dataset covering nine programming languages and 7.9 million pull requests from over 1,000 GitHub repositories, designed to support tasks like review comment generation. Yu et al. \cite{yu2024fine} further refined annotation quality by employing ChatGPT for automated data labeling and validation.

For security-related code review, Yu et al. \cite{yu2023security} curated a dataset from OpenStack and Qt projects. Using a keyword-based extraction and manual annotation, they identified 614 security-related review comments, categorizing them into 15 security defect types. The dataset was aligned with CWE taxonomy to enhance validity.
\section{Conclusion and Future Work}

In this paper, we present \textbf{SeRe}, a security-related code review dataset, constructed through an automated approach based on an active learning-powered ensemble classifier. Comprising 6,732 security-related reviews spanning diverse programming languages, SeRe aligns closely with real-world review distributions, providing a high-quality benchmark for researchers and practitioners. In addition, we investigate the effectiveness of the state-of-the-art code review comment generation approaches, evaluating their performance on security-related contexts. Our findings reveal the strengths and limitations, highlighting the challenges posed by security-related code reviews.

In the future, we will expand the dataset to cover more security concerns and projects. Furthermore, we will investigate the applicability of our data construction methodology to other types of code reviews, such as refactoring suggestions, to assess its broader utility. Additionally, we will develop advanced approaches for code review generation by leveraging our dataset to improve both accuracy and relevance.

\begin{acks}
The authors would like to say thanks to anonymous reviewers for their insightful comments and suggestions. This work was partially supported by the National Natural Science Foundation of China (62232003, 62172037 and 62172214), and CCF-Huawei Populus Grove Fund.
\end{acks}

\balance
\bibliographystyle{ACM-Reference-Format}
\bibliography{references}
\end{document}